\documentclass[useAMS,usenatbib]{mn2e}
\usepackage{psfig}
\def\mch{M$\rm^{c}$Hardy\,}
\def\etal{et al.~}

\def\xmm{{\em XMM-Newton }}

\def\xte{{\em RXTE }}
\def\ecs{ergs cm$^{-2}$ s$^{-1}$~}
\def\mcg6{MCG--6-30-15}
\def\c{{Cyg X-1}}
\def\msun{$M_{\odot}$}
\def\ltsim{\mathrel{\hbox{\rlap{\hbox{\lower4pt\hbox{$\sim$}}}\hbox{$<$}}}}
\def\gtsim{\mathrel{\hbox{\rlap{\hbox{\lower4pt\hbox{$\sim$}}}\hbox{$>$}}}}
\def\mdot{{$\dot{m}$}~}

\begin{document}

\title[Long Timescale X-Ray Variability of MCG--6-30-15]
{MCG--6-30-15: Long Timescale X-Ray Variability, Black Hole Mass
and AGN High States}

\author[M$\rm^{c}$Hardy, I.M., \etal]
{I M M$\rm^{c}$Hardy,$^{1}$ K F  Gunn,$^{1}$ P Uttley,$^{1}$ M R Goad,$^{1}$
\\
$^{1}$ Department of Physics and Astronomy, The University,
Southampton SO17 1BJ
}

\date{Accepted for publication in Mon. Not. R. Astron. Soc.}
\pagerange{\pageref{firstpage}--\pageref{lastpage}}
\pubyear{2005}

\label{firstpage}

\maketitle

\begin{abstract}
We present a detailed study of the long-timescale X-ray variability of
the Narrow Line Seyfert 1 Galaxy (NLS1) MCG--6-30-15, based on eight
years of frequent monitoring observations with the {\it Rossi X-ray
Timing Explorer}.  When combined with the published short timescale
{\it XMM-Newton} observations, we derive the powerspectral density
(PSD) covering 6 decades of frequency from $\sim10^{-8}$ to
$\sim10^{-2}$Hz. As with NGC4051, another NLS1, we find that the PSD
of MCG--6-30-15 is a close analogue of the PSD of a Galactic Black
Hole X-ray binary system (GBH) in a `high' rather than a `low'
state. As with NGC4051 and the GBH Cygnus~X-1 in its high state, a
smoothly bending model is a better fit to the PSD of MCG--6-30-15,
giving a derived break frequency of $7.6^{+10}_{-3}
\times 10^{-5}$ Hz.  Assuming linear scaling of break frequency with
black hole mass, we estimate the black hole mass in
\mcg6 to be $\sim2.9^{+1.8}_{-1.6} \times 10^{6}$\msun.

Although, in the X-ray band, it is one of the best observed Seyfert
galaxies, there has as yet been no accurate determination of the mass
of the black hole in MCG--6-30-15.  Here we present a mass
determination using the velocity dispersion ($M_{\rm BH}-\sigma_{*}$) technique
and compare it with estimates based on the width of the $\rm
H \alpha$ line.  Depending on the calibration relationship assumed
for the $M_{\rm BH}-\sigma_{*}$ relationship,  we derive a
mass between 3.6 and 6 $\times 10^{6}$\msun, consistent with the
mass derived from the PSD.

Using the newly derived mass and break timescale, and revised
reverberation masses for other AGN from \cite{peterson04}, we update
the black hole mass/break timescale diagram.  The observations are
still generally consistent with narrow line Seyfert 1 galaxies having
shorter break timescales, for a given mass, than broad line AGN,
probably reflecting a higher accretion rate.  However the revised,
generally higher, masses (but unchanged break timescales) are also
consistent with perhaps all of the X-ray bright AGN studied so far
being high state objects.  This result may simply be a selection
effect, based on their selection from high-flux X-ray all sky
catalogues, and their consequent typically high X-ray/radio ratios,
which indicate high state systems.

\end{abstract}


\section{Introduction}
It is now reasonably well established that the X-ray powerspectral
densities (PSDs) of AGN are broadly similar to those of galactic
black hole X-ray binary systems (GBHs)
\citep{mch88,ednan99,uttley02,markowitz03a,mch04}. 
The PSDs are described by powerlaws of the form
$P(\nu) \propto \nu^{\alpha}$, where $P(\nu)$ is the power at
frequency $\nu$, but $\alpha$ varies with frequency (see Section 3.2.1 
for details). At high frequencies the
PSDs of both AGN and GBHs are steep (slope, $\alpha, \sim-2$) but below a
break frequency, $\nu_{B}$, typically at a few Hz for GBHs, 
they flatten to a slope of $\alpha \sim-1$.
To first order, the break timescale scales linearly with the black
hole mass but there is still considerable uncertainty in the
mass/break timescale relationship and hence in our understanding
of the physical similarities between AGN and GBHs.

The uncertainty arises largely because GBHs occur in 
several distinct states
\cite[e.g. see][]{mcclintock03}. Most commonly they are found in the
so-called `low' state where their X-ray fluxes are low and their
medium energy (2-10 keV) X-ray spectra are hard. The second most
common state is the `high' state where their fluxes are high and their
2-10 keV spectra are soft. The PSDs of GBHs in these two states
are quite different. In the low state,
there is a second break, about a decade below the high frequency break
\citep{nowak99}.  Below the lower frequency break the PSD flattens
further to a slope of zero. In the high state there is no second break
and the PSD continues with slope $\sim-1$ to very low frequencies
\citep{cui97a}. In addition the break from slope $\sim-2$ to slope
$\sim -1$ occurs at a higher frequency in the high state than in
the low state ($\sim 15$Hz cf $\sim3$Hz in the best studied GBH Cyg
X-1). Before comparing AGN with GBHs it is therefore important to
know whether the AGN is a high or low state system.

Based on their 2-10 keV X-ray spectra, it has
generally been assumed that AGN are the equivalent of low state
GBHs. However, although the high frequency ($\nu >10^{-6}$Hz)
parts of many AGN PSDs are now reasonably well determined, the
lower frequencies are, in general, not well determined and so it
is not usually possible to be sure whether AGN are the analogues
of low or high state GBHs. In general it is not possible to
determine whether there is a second, lower frequency, break or
not. The difficulty in determining the low frequency shape of AGN
PSDs has been in obtaining well sampled lightcurves stretching
over sufficiently long timescales. For an assumed linear scaling
of break timescale with mass, we require well sampled lightcurves
of a few years duration to detect the second, lower, break in a
black hole of mass $\sim10^{6}$ \msun.

Prior to the launch of \xte in November 1995 it was not possible to
obtain long timescale lightcurves of sufficient quality but, since
early 1996 we
\citep{mchxte98,uttley02,mch04}, and others \citep{ednan99,markowitz03a},
have been monitoring a small sample of AGN, typically 
observing each AGN once every 2
days, and so are able to determine the shape of the low
frequency PSD to high precision. Using these observations we
\citep{mch04} recently showed that the PSD of the narrow line
Seyfert 1 (NLS1) galaxy NGC~4051 was, in fact, identical to that of a
high, rather than low state GBH and so provided the first definite
confirmation of a high state AGN. \cite{markowitz03a} plotted the black hole
mass against the timescales associated with well measured PSD breaks,
some from slope $\sim-2$ to $\sim-1$ and some from slope $\sim-1$ to
$\sim0$. They claimed that there was a linear relationship between the
two parameters.  With a somewhat larger sample we \citep{mch04}
plotted the black hole mass against the timescale associated with the
specific PSD break from slope $\sim-2$ to $\sim-1$ for a sample of
AGN. We found that the best fit to all of the AGN together
had a slope flatter than unity and did not extrapolate to either the
high or low state break timescales of Cyg X-1. However we noted that
the break timescales associated with broad line Seyfert 1 galaxies
were, for a given black hole mass, generally longer than those for
those objects usually classed as
NLS1s. NLS1s are probably not a distinct class of object but, more likely,
just lie at one end of a spectrum of AGN properties, characterised perhaps
by a higher accretion rate. However taking the crude broad/narrow line 
distinction which is commonly used in the literature, we note that
a linear BH-mass/break-timescale relationship fits broad line
AGN and Cyg X-1 in the low state, and a displaced linear relationship
fits the NLS1s and Cyg X-1 in the high state. We
therefore suggested that there is no fixed break timescale/mass
relationship which fits all AGN but that the break timescale/mass
relationship may actually vary in response to one or more other
factors, which might be accretion rate or black hole spin.

In order to test and clarify the above hypothesis, it is necessary to
place AGN on the break timescale/mass plane with high precision and to
determine whether they are high or low state analogues. There are very
few AGN for which the X-ray observations are extensive enough to make
the latter determination possible and a particularly important object
is therefore the Seyfert galaxy \mcg6.  Although \mcg6 is not always
described as an NLS1, it has many of the properties commonly
associated with NLS1s, ie rapid X-ray variability and relatively narrow
permitted emission lines (FWHM 1700 km $\rm
s^{-1}$ \citep{pineda80} c.f. the 2000 km $\rm s^{-1}$ limit
which is usually quoted for NLS1s). In this paper we therefore class \mcg6
as a NLS1, like NGC4051.

\begin{figure*}
\psfig{figure=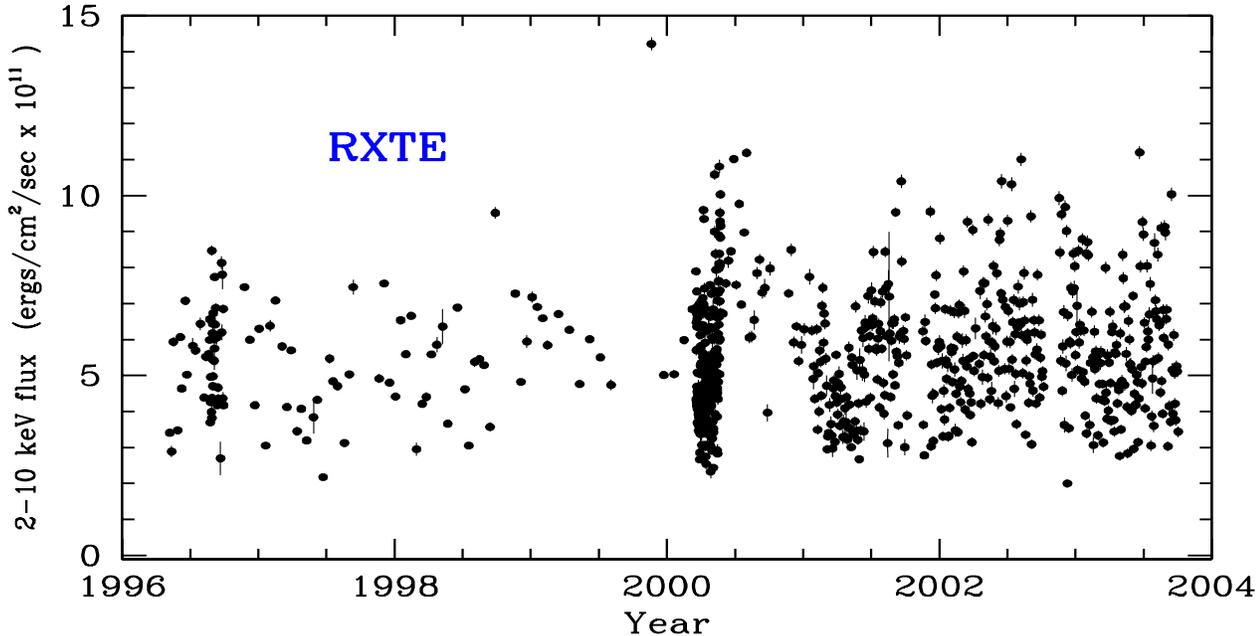,width=6.8in,height=4in,angle=0}
\caption{Long Term {\it RXTE} 2-10 keV lightcurve of MCG--6-30-15.
Each data point represents an observation of $\sim1$~ks.}
\label{fig:xtelc}
\end{figure*}

\mcg6 is one of the best studied X-ray bright Seyfert galaxies. It
was the galaxy in which the first detection of a relativistically
broadened X-ray iron line was made, providing direct evidence for
the existence of a massive black hole \citep{tanaka95}. We have
monitored it extensively with the Rossi X-ray Timing Explorer
({\it RXTE}) and have presented its long timescale PSD
\citep{uttley02}, based on 3 years of observations (1996-1999). 
From that early, limited, dataset \cite{uttley02} were able to show that
a simple unbroken powerlaw was not a good fit to the PSD (at greater
than 99 percent rejection confidence) but that a broken powerlaw did provide
a good fit (rejection confidence 33 percent). However it was not possible to
determine, with any accuracy, the low frequency PSD slope. If they
assumed a powerlaw of slope -1, Uttley et al. were able to derive a
break frequency in the PSD at $\sim$few $\times10^{-5}$Hz and to
estimate the high frequency powerlaw slope ($\sim -2$). Thus Uttley et al. were
unable to distinguish a high from a low state system, or measure the
break timescale with high precision, although they preferred the high
state interpretation as a low state interpretation implied a
luminosity close to the Eddington limit. \cite{vfn03} have
subsequently published a detailed study of the short timescale X-ray
variability of MCG--6-30-15 using observations from
{\it XMM-Newton}. These observations are insufficient, on their own, to
determine the low frequency PSD slope but, by also assuming a low
frequency PSD slope of -1, they have determined the break frequency
to be $\sim1 \times 10^{-4}$Hz.

In this paper (Section~\ref{sec:xobs}) we present our full \xte
observations, of 8 years duration, and consisting of over 800
separate observations, compared to 100 in \cite{uttley02}. In
Section~\ref{sec:psd} we combine these \xte observations with the
published {\it XMM-Newton} observations and derive a PSD of excellent quality
covering over 6 decades of frequency from $<10^{-8}$ to
$>10^{-2}$Hz. We determine accurately the shape of the PSD below
the $\sim10^{-4}$Hz break and we thereby show that \mcg6 is, like
NGC4051 \citep{mch04}, the analogue of a high state GBH. Using our
improved determination of the low frequency PSD shape, we slightly
refine the break timescale.

The most widely accepted method of determining
black hole masses in Seyfert galaxies is that of
reverberation mapping \cite[e.g.][]{peterson01}
but this technique has not yet been applied to \mcg6.
The best current estimate of the BH mass in \mcg6 is $\sim1
\times 10^{6} \rm M_{\odot}$ \citep{uttley02}. 

That mass is based on an estimated bulge mass of $3 \times 10^{9} \rm
M_{\odot}$ \citep{reynolds00} and on the correlation between
black hole mass and galactic bulge mass presented by \cite{wandel99},
where he claims that Seyfert galaxies have a lower black hole mass,
for a given bulge mass, than was claimed in the original relationship
of \cite{magorrian98}. In a later paper \cite{wandel02} revises the
black hole mass/bulge ratio to 0.0015 and shows that Seyfert galaxies fit 
the same relationship as normal galaxies. Using the revised
ratio, and the estimated bulge mass from \cite{reynolds00},
the revised black hole mass would be $4.5 \times 10^{6} \rm M_{\odot}$.

An upper limit to the black hole mass of $10^{7}$\msun has been derived by
\cite{morales02}, using a method based on balancing the radiative and
gravitational forces acting on outflowing warm absorber clouds.
However it has recently become clear that the successful
technique of determining black hole masses from the central
velocity dispersion of the galaxy bulge can be applied to active,
as well as quiescent, galaxies by observing the width of the Ca{\sc ii}
triplet in the far red, where the AGN continuum is not dominant
\citep{ferrarese01}.
In Section~\ref{sec:veldisp} we report the determination of the
black hole mass in MCG--6-30-15 from central velocity dispersion
measurements. For comparison (Section~\ref{sec:linewidth}) we also
estimate the BH mass from the width of the $\rm H \beta$ emission
line and the continuum flux at 5100\AA, using empirical relations
from \cite{kaspi00}. With greater error, we also estimate
(Section~\ref{sec:photoion}) the BH mass using the
`photoionisation' method \citep{wandeletal99}.

In Section~\ref{sec:disc} we discuss the implications of our
results for the comparison of AGN and GBHs.

\section{X-RAY OBSERVATIONS AND DATA REDUCTION}
\label{sec:xobs}

\subsection{{\it RXTE}}
\label{sec:xte}
The \xte observations discussed here consist of our own
monitoring observations, which have continued since 1996, and two
`long looks' in August 1997 and July 1999 which we have obtained from the 
\xte archive. The monitoring observations cover timescales from less than a
day to $\sim$few years and the long looks cover timescales from $\sim$days to 
$\sim$minutes.

\subsubsection{Monitoring Observations}

The \xte monitoring observations, of typically $\sim$1~ks duration, were made
and analysed in exactly the same way as for NGC4051 \citep{mch04}.
Prior to 2000, we used a quasi-logarithmic sampling pattern, covering
all timescales, but in order to improve the S/N
on our resulting PSDs, we then increased our coverage. From 2000
onwards, our typical observation frequency has been once every two
days. In addition, to sample properly the higher frequencies, we
observed once every 6 hours for 2 months in 2000.
Each observation, of $\sim1$ksec duration, contributes one flux point
to the monitoring lightcurve. We do not attempt to split the
monitoring data into higher time resolution bins.

The observations were made with the Proportional Counter
Array (PCA) which consists of 5 Xenon-filled proportional counter
units (PCUs). We extracted the Xenon 1 (top) layer data from all
PCUs that were switched on during the observation as layer 1
provides the highest S/N for photons in the energy range 2-20 keV
where the flux from AGN is strongest.  We used FTOOLS v4.2 for the
reduction of the PCA data. We used standard `good time' data
selection criteria, i.e. target elevation $>10^{\circ}$, pointing
offset $<0.01^{\circ}$, time since passage through the South
Atlantic Anomaly of $>30$min and standard threshold for electron
contamination.  We calculated the model background in the PCA with
the tool PCABACKEST v2.1 using the L7 model for faint sources.
PCA response matrices were calculated individually for each
observation using PCARSP V2.37, taking into account temporal
variations of the detector gain and the changing number of
detectors used.  Fluxes in the 2-10 keV band were then determined
using XSPEC, fitting a simple powerlaw with variable slope but
with absorption fixed at the Galactic level of $4.06 \times
10^{20}$ cm$^{-2}$ \citep{elvis89}. 
The errors in the flux are scaled
directly from the observed errors in the measured count rate.

As with NGC4051, we produce a lightcurve in flux units, rather than
raw observed counts/sec, so that we may use together data from periods
when the number, and gain, of the PCUs changes.  The resulting
lightcurve, from 1996 to 2004, is presented in Fig.~\ref{fig:xtelc}.

From the monitoring observations we make three lightcurves.  The 6-hr
sampled lightcurve consists of the 2 months of observations 4
times per day. The 2-d sampled lightcurve consists of the period 
from 2000 onwards where sampling is once every 2 days. The total
lightcurve consists of all of the observations since 1996, but when
used in the PSD analysis it is always heavily binned to 28 or 56-d
resolution. 

\begin{figure}
\psfig{figure=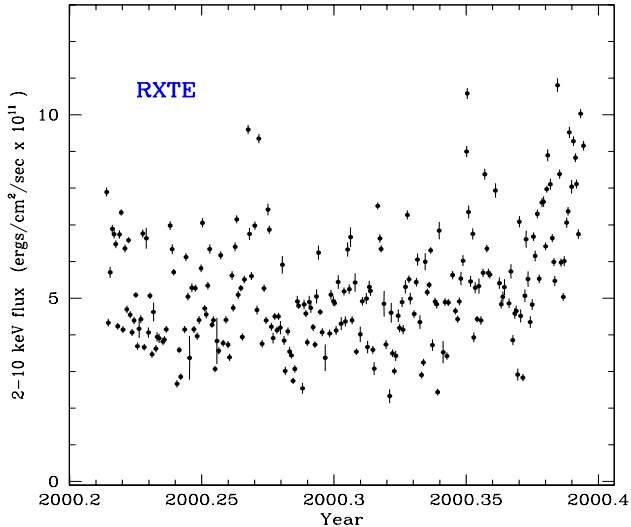,width=3.4in,height=3in,angle=0}
\caption{ The 2-10 keV {\it RXTE} lightcurve of MCG--6-30-15
covering the two-month period of four observations per day.
Each data point again represents an observation of $\sim1$~ks.}
\label{fig:intensive_lc}
\end{figure}

\subsubsection{Long Looks}

There have been two long looks with \xte, one in August 1997 and the
other in July 1999.  The total duration of each observation was about
9 days, consisting of continuous segments of $\sim3$ksec, separated by
gaps of approximately equal length caused by earth occultation. The
August 1997 observations have been published by \cite{lee00} who
have carried out a preliminary analysis of their variability
properties. \cite{lee00} claim a break in the powerspectrum at $\sim4-5
\times 10^{-6}$Hz. \cite{nowakchiang00} have analysed the same
dataset, together with ASCA observations,
and claim that the resultant PSD resembles that of a GBH in a
low state. The claim that the PSD is flat below $10^{-5}$Hz,
has a slope of approximately -1 from $10^{-5}$Hz to $\sim10^{-4}$Hz
and, above $\sim10^{-4}$Hz it steepens to a slope of -2. Neither
Lee et al. or Nowak and Chiang carry out simulations to
determine the validity of their PSDs. \cite{uttley02} do carry out
simulations and conclude that that although there is a break in the
PSD of \mcg6, the model of Nowak and Chiang can be rejected at
99 percent confidence and that there is significant power in the PSD
below $10^{-5}$Hz. 
A time-series analysis of the July 1999
observations has not yet been published.

\begin{figure}
\psfig{figure=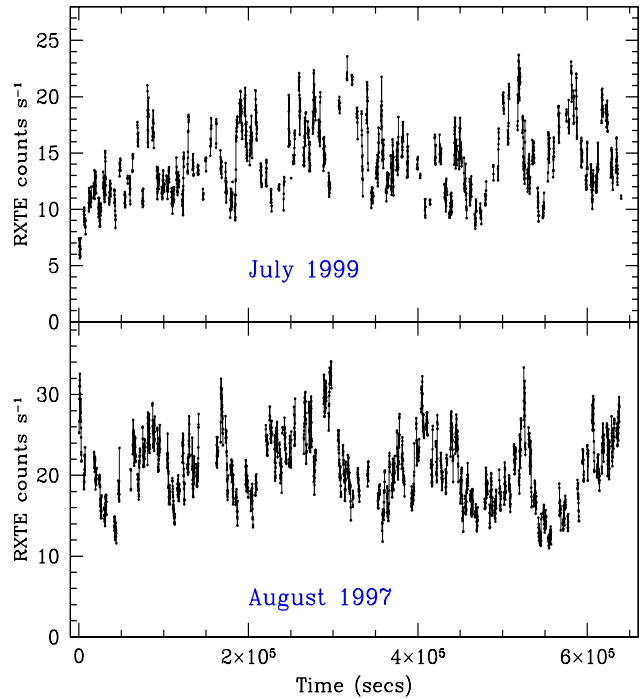,width=9cm,height=10cm,angle=0}
\caption{{\em RXTE} background-subtracted lightcurve of MCG--6-30-15 in
the 2-10 keV energy band, with 128s time bins. 
The lightcurves are shown as histogram segments. Gaps are left where
there are no data. Bottom panel is the
August 1997 observation and top panel is the July 1999 observation.
In both cases the time is measured since the start of the observation. }
\label{fig:longlook}
\end{figure}

The two long looks were analysed in exactly the same way as the
monitoring observations, except that we retained 16s time resolution.
As the July 1999 lightcurve has not previously been published, we
include it here (Fig.~\ref{fig:longlook}) and, for the convenience of
the reader, we also include the August 1997 lightcurve which has previously
been published by Lee et al.

\subsection{{\em XMM-Newton}}
\label{sec:xmm}
\begin{figure}
\psfig{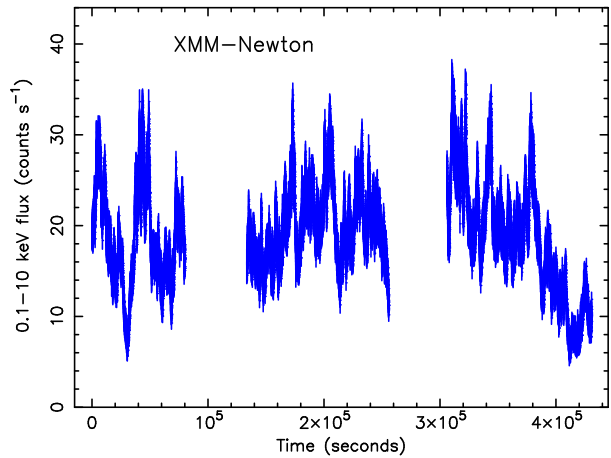}
\caption{{\em XMM-Newton} background-subtracted lightcurve of MCG--6-30-15 in
the 0.1-10 keV energy band, with 20s time bins, using data from
the PN CCDs.}
\label{fig:xmmlc}
\end{figure}

The {\it XMM-Newton} timing observations of \mcg6 have been discussed
extensively by \cite{vfn03} to which we refer readers for a detailed
discussion.  For the convenience of readers, we reproduce their
lightcurve here (Fig~\ref{fig:xmmlc}).

\section{PSD Determination}
\label{sec:psd}

The method used to determine the PSD is the same Monte Carlo
simulation-based modelling technique \citep{uttley02}, {\sc psresp},
which we employed in the analysis of the combined {\it XMM-Newton} and
\xte observations of NGC4051
\citep{mch04}. This method is able to take account of non-uniform
sampling and gaps in the data.

\subsection{Long Timescale PSD}
A variety of datasets are available for determination of the overall
PSD but the \xte monitoring observations provide the only information
on timescales longer than $\sim$days. We therefore begin by fitting a
simple power law to the PSD from the combined 6-hr, 2-d and total \xte
lightcurves. We retain the intrinsic 6-hr and 2-d resolution for the
first two lightcurves but bin the total lightcurve up to 28-d
resolution. 

\begin{figure}
\psfig{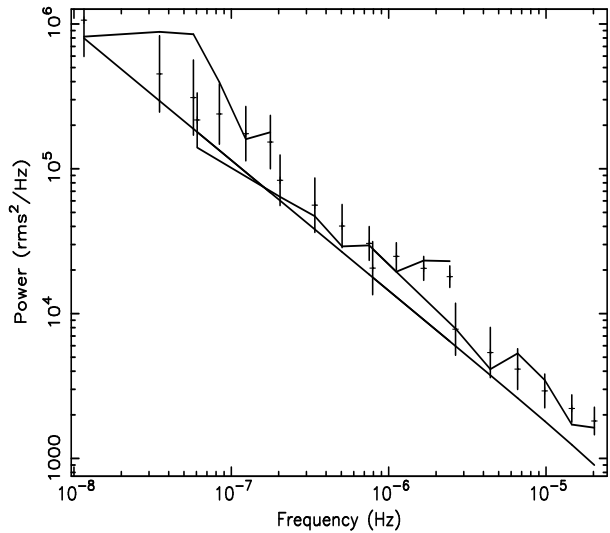}
\caption{Long timescale \xte PSD of MCG--6-30-15. The PSD is well
fitted (P=68 percent) by a simple powerlaw of slope $0.9\pm0.25$ over three
decades. The underlying undistorted model PSD is shown by the smooth
continuous line. The distorted model for each of the three datasets, 
together with its errors, is shown by the individual points. 
The observed dirty PSD is given by the jagged lines.
}
\label{fig:xtelongpsd}
\end{figure}

The resulting PSD is well fitted (fit probability = 68 percent)
by a slope of $0.9\pm0.15$ (90 percent confidence intervals are used
throughout this paper). 
In \cite{uttley02} we describe how we determine the goodness of fit
from the simulations. A fit probability of 68 percent means that the model
is rejected at 32 percent confidence.
The fit is shown in Fig.~\ref{fig:xtelongpsd}.
For a GBH in the low-hard state, we do not expect the
region of slope $\sim-1$ to extend for more than one or 1.5
decades, before flattening to a slope near zero.
Although this fit would not be very sensitive to breaks near to
the end of the frequency spectral range covered, nonetheless the
fit to a simple power law, of slope close to $-1$ over approximately
three decades, from $\sim10^{-8}$ to $\sim10^{-5}$ Hz, suggests
that \mcg6 is the analogue of an XRB in a high-soft state rather
than in a low-hard state.

\subsection{Combined Long and Short Timescale PSD}

\subsubsection{\xte monitoring observations and {\it XMM-Newton} 4-10 keV
observations}

In order to determine properly the overall long and short timescale
PSD, and hence measure any break frequency and high frequency PSD
slope, and refine measurements of the low frequency PSD slope, we must
combine the \xte monitoring observations with observations which
sample shorter timescales.

Both the \xte long looks and the {\it XMM-Newton} long observations
provide information on shorter timescales. The
\xte long looks provide a good determination of the PSD on timescales
shorter than the length of the individual segments
($\sim3000$s). However on timescales between $\sim3000$s and $\sim$few
days, the PSD is distorted by the many gaps in the datasets. Our
{\sc psresp} analysis software is able to cope with those gaps,
but they do give the software considerable work to do as the dirty PSD
differs a good deal from any likely underlying true PSD. Thus the
errors are larger than if there are fewer gaps. The {\it XMM-Newton}
data are continuous and so the {\it XMM-Newton} PSD suffers negligibly
from distortion. However the sensitivity of {\it XMM-Newton} is less
than that of \xte in the 2-10 keV range, which is the only energy band
in which the long timescale PSD is determined. Nonetheless, the long
{\it XMM-Newton} observations of \mcg6, which total approximately
three times as long as those of NGC4051, do allow a reasonable
determination of the high frequency PSD.

As discussed in \cite{mch04}, the mean photon energy of the 4-10 keV \xmm
band is approximately the same as that of the \xte 2-10 keV band.
In order to derive a PSD of uniform energy at all frequencies
we have therefore included the 4-10 keV {\it XMM-Newton} PSD with the \xte
monitoring datasets described above in our PSD simulation process.
The high-soft state PSD of Cyg X-1 is better described by a bending
powerlaw model than a sharply broken powerlaw model, and the PSD of
NGC4051 is also slightly better fit by the bending powerlaw model
\[ P(\nu) = A \nu^{\alpha_{L}} \, \left( 1 + 
\left(\frac{\nu}{\nu_{b}} \right)^{(-\alpha_{H} + \alpha_{L})}
\right)^{-1}\, \]
where $\alpha_{L}$ and $\alpha_{H}$ are the low and high frequency
powerlaw slopes respectively and $\nu_{b}$ is the bend, or break,
frequency. As stated in the Introduction,
here we use the convention that a slope decreasing
towards high frequencies will be defined by $P(\nu) \propto
\nu^{\alpha}$ where $\alpha$ is a negative number, eg $\alpha=-2$. 
\footnote{This convention is used throughout the Tables and text of 
\cite{mch04}. Note that when we defined the formula
for a bending powerlaw in \cite{mch04}(top of second column, page 788),
we forgot to make the formula
consistent with the rest of the text and Tables and so, only in that
one formula, a slope decreasing towards high frequencies 
is defined by $P(\nu) \propto
\nu^{-\alpha}$ where $\alpha$ is a positive number, eg $\alpha=+2$.}

\begin{figure}
\psfig{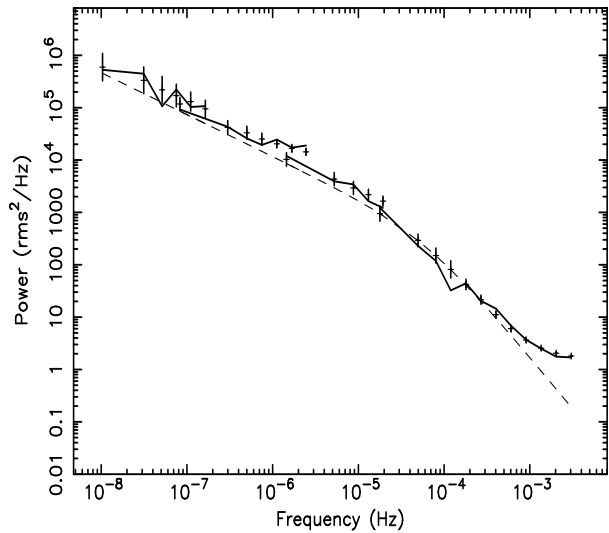}
\caption{Combined 2-10 keV \xte and  4-10 keV {\it XMM-Newton} PSD of MCG--6-30-15.
The lines and datapoints are as described in Fig.~\ref{fig:xtelongpsd}.
}
\label{fig:xte410psd}
\end{figure}

\begin{figure}
\psfig{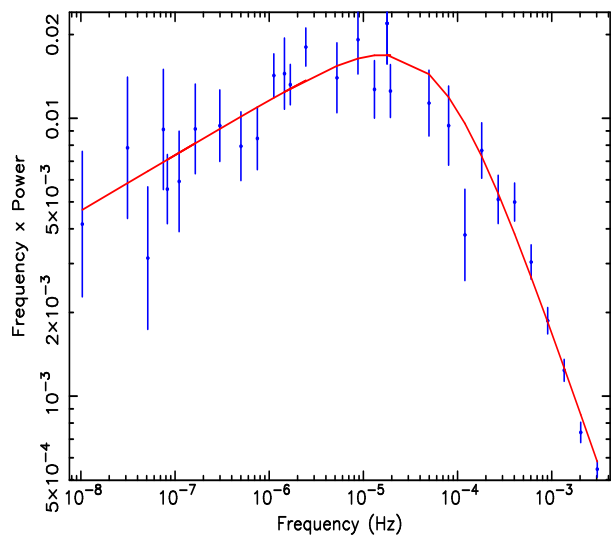}
\caption{Unfolded combined 2-10 keV \xte and  4-10 keV {\it
XMM-Newton} PSD of MCG--6-30-15. Here the errors have been translated
onto the datapoints and the continous line represents the best-fit
underlying PSD. Note that here we plot frequency $\times$Power. See text for details.
}
\label{fig:unfolded}
\end{figure}

We fit this model to the combined \xte and
{\it XMM-Newton} observations and the result is shown in
Figs.~\ref{fig:xte410psd} and ~\ref{fig:unfolded}.
In Fig.~\ref{fig:xte410psd} we plot the log of power vs. the log of
frequency, as in Fig.~\ref{fig:xtelongpsd}.
In both of these figures the observed dirty PSDs  from the various
constituent lightcurves are given by the jagged lines. The model,
distorted by the effects of sampling, red noise leak and aliasing, is
given by the points with errorbars, and the underlying, undistorted,
model is given by the smooth dashed line.
In Fig.~\ref{fig:unfolded} we unfold the observed dirty PSD from the
distorting effect of the sampling in order to produce the closest
approximation that we can to the true underlying PSD. 
Thus displacements of the observed PSD from the model distorted PSD are
translated into displacements from the best-fit underlying model PSD.
The technique is identical to that used to deconvolve energy spectra
from count rate (ie instrumental) spectra in standard X-ray spectral fitting.
In this case the error is translated onto the observed datapoints and the
underlying best-fit model is shown as a continuous line.
As in standard X-ray spectral fitting, the position of the datapoints
does depend on the shape of the assumed, best-fitting, model.
Note that in Fig.~\ref{fig:unfolded} we plot frequency $\times$power
so that a horizontal line would represent $Power(\nu) \propto
\nu^{-1}$ and equal power in each decade. Similar plots for NGC4051,
NGC3516 and Cyg X-1 are shown in Fig.18 of \cite{mch04}.

\begin{figure}
\psfig{figure=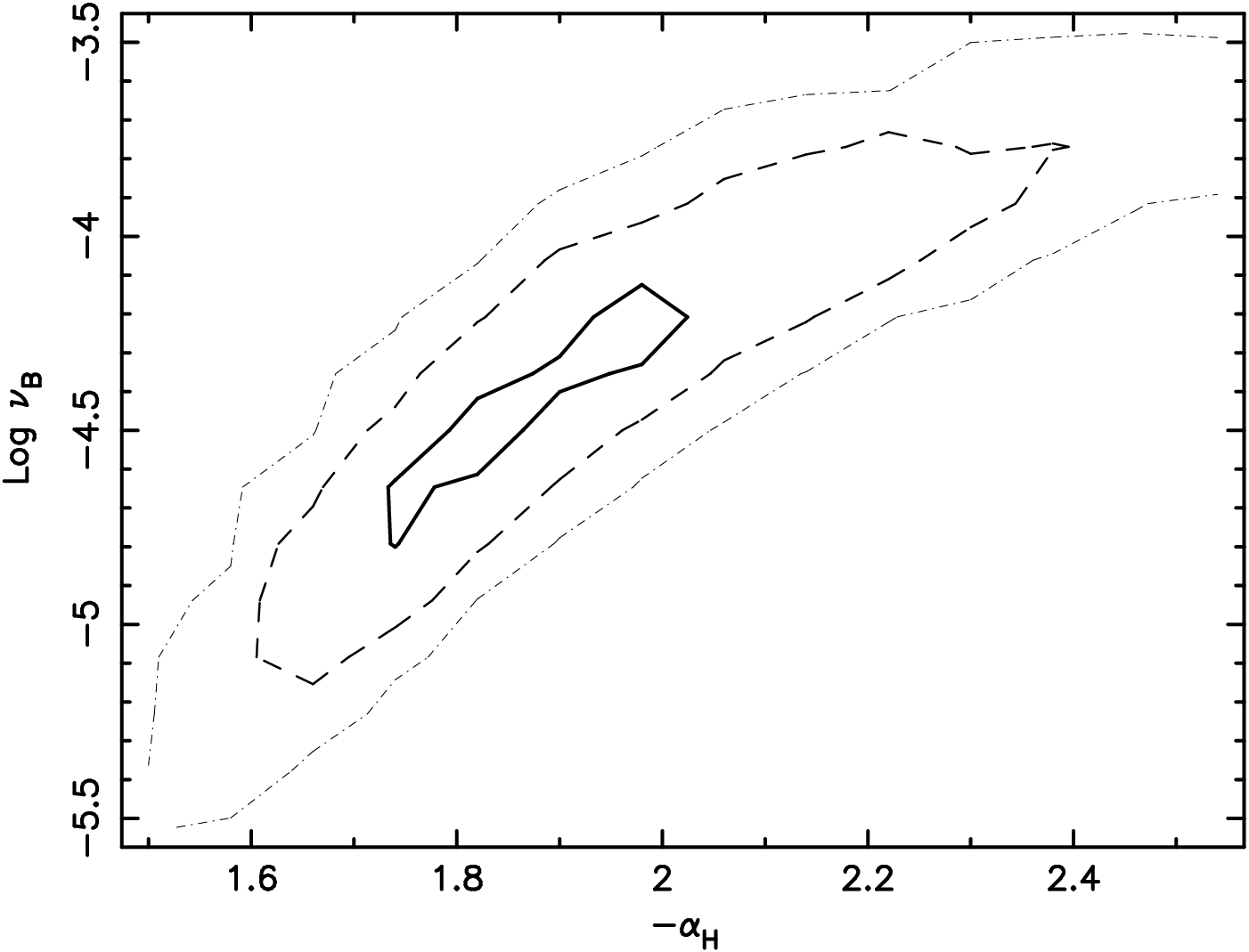,width=8cm,height=7cm,angle=270}
\caption{68 percent, 90 percent and 99 percent confidence contours for 
high frequency slope, $\alpha_{H}$, and break frequency, $\nu_{B}$, for bending
powerlaw fit to the combined {\it RXTE} and {\em XMM-Newton} 4-10 keV
PSD shown in Fig.~\ref{fig:xte410psd}. Note we plot $-\alpha_{H}$.}
\label{fig:con13_xte410}
\end{figure}

The combined PSD is well fitted (P=45 percent) by this model with low
frequency slope $\alpha_{L}=-0.8^{+0.4}_{-0.16}$, high frequency slope
$\alpha_{H}=-1.98^{+0.32}_{-0.40}$ and break frequency
$\nu_{B}=6.0^{+10}_{-5} \times 10^{-5}$Hz.  A sharply broken powerlaw,
although tolerable, is not such a good fit (P=14 percent) but the best fit
parameters are similar to those of the smoothly bending model
($\alpha_{L}=-0.8$, $\alpha_{H}=-1.74$ and $\nu_{B}=3.2 \times
10^{-5}$Hz).
As is often the case, the break frequency is slightly lower for the
sharply bending model than for the smoothly bending model but, in this 
case, the errors are not well determined.
(Note that our software
performs a simple grid search so the values given here are those of
the minimum point on the grid and so may differ very slightly from the
optimum values which may lie slightly off any particular grid point.)
The confidence contours for these parameters are shown in
Figs.~\ref{fig:con13_xte410}, \ref{fig:con14_xte410} and
\ref{fig:con34_xte410}. 

The break frequency for \c, assuming a high state, bending powerlaw
model, is $22.9 \pm1.5$Hz \citep{mch04} (or $13.9 \pm 0.8$Hz for a
sharply breaking high state model).  Assuming linear scaling of black
hole mass with frequency, and a black hole mass of 10\msun for \c \,
\citep{herrero95}, we derive a black hole mass for \mcg6 of
$3.6^{+18}_{-2} \times 10^{6}$\msun.

\begin{figure}
\psfig{figure=con14_410xmmpsd_chop2_2_neg.ps,width=8cm,height=7cm,angle=270}
\caption{68 percent, 90 percent and 99 percent confidence contours for low frequency
slope, $\alpha_{L}$,
and break frequency, $\nu_{B}$, for bending powerlaw fit to the combined
{\it RXTE} and {\em XMM-Newton} 4-10 keV PSD shown in Fig.~\ref{fig:xte410psd}.
Note we plot $-\alpha_{L}$}
\label{fig:con14_xte410}
\end{figure}
 
\begin{figure}
\psfig{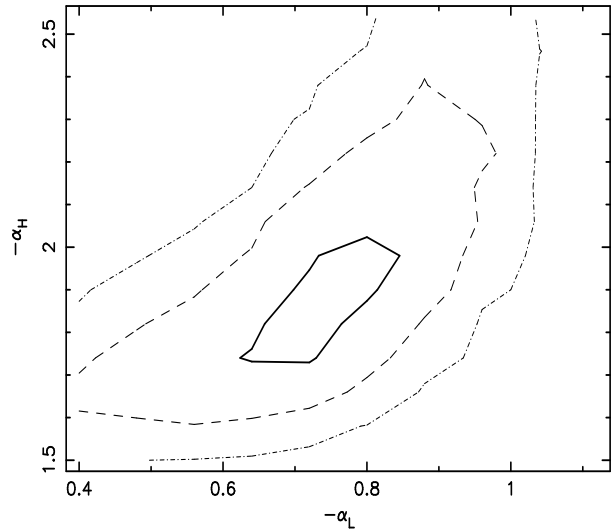}
\caption{68 percent, 90 percent and 99 percent confidence contours for low and high 
frequency slopes, $\alpha_{L}$ and $\alpha_{H}$ respectively for
bending powerlaw fit to the combined {\it RXTE} and {\em XMM-Newton}
4-10 keV PSD shown in Fig.~\ref{fig:xte410psd}. Note we again plot
$-\alpha_{L}$ and $-\alpha_{H}$.
}
\label{fig:con34_xte410}
\end{figure}

We note that, although consistent within the errors with the break
frequency derived by \cite{vfn03} using purely {\it XMM-Newton}
observations, our break frequency is slightly lower. The reason for
the difference is that Vaughan et al. were unable to measure the PSD
slope below the break and so assumed a value of -1. However our data
show that the slope is actually slightly flatter, which leads to a
lower break frequency.

\subsubsection{\xte monitoring observations and low energy {\it XMM-Newton} observations}

In \cite{mch04} we refined our determination of the break frequency 
in NGC4051 by 
combining \xte monitoring observations with continuous {\it XMM-Newton}
observations in the
0.1-2 keV band, where the break is better defined. We assumed that the 
break frequency was independent of energy. We have carried out 
the same procedure here (again taking proper account of the different
PSD normalisations between the 0.1-2 and 2-10 keV bands)
and find that
$\alpha_{L}=-0.8^{+0.2}_{-0.1}$,
$\alpha_{H}=-2.5^{+0.3}_{-0.4}$ and 
$\nu_{B}=7.6^{+10}_{-3} \times 10^{-5}$Hz.
The fit probability is 67 percent. 
Apart from the steeper high energy slope which is well known at low
energies \citep{vfn03,mch04}, the other fit parameters are very similar 
to those obtained when using the \xte and {\it XMM-Newton} 
4-10 keV observations.
Thus within the errors, we find no evidence for a change of break
frequency with photon energy. The implied black hole mass is more
tightly constrained by the smaller errors on the break frequency and
is $2.9^{+1.8}_{-1.6} \times 10^{6}$\msun.

We fitted a sharply breaking powerlaw model to the 
combined \xte and \xmm 0.1-2 keV data. As with the combined \xte and
4-10 keV \xmm data,
the fit parameters are almost exactly the same as for the bending
powerlaw model, but the fit probability is worse (19 percent).
The fact that a bending powerlaw is a better fit 
than a sharply bending powerlaw to the PSD of \mcg6,
as it is to the PSD of \c in its high state, strengthens our
conclusion that \mcg6 is the analogue of a GBH in the high, rather
than low, state.

\subsubsection{\xte monitoring observations and \xte long looks}

We have also determined the overall PSD shape by including the \xte
long looks binned up to 128s time resolution
with the \xte monitoring observations, and also including a 
very high frequency PSD ($>10^{-3}$Hz) made from the individual
segments of the \xte long looks. The fit parameters are approximately
the same as in the combined \xte and {\it XMM-Newton} fit 
($\alpha_{L}=-0.8^{+0.5}_{-0.15}$,
 $\alpha_{H}=-2.1^{+0.4}_{-0.6}$ and
$\nu_{B}=5.0^{+8}_{-4} \times 10^{-5}$Hz).
The fit is formally good (P=74 percent) but the errors are rather large
because of the difficulty of coping with 
the many gaps in the \xte long look lightcurves.

\subsection{Search for a second, lower frequency, break}

\begin{figure}
\psfig{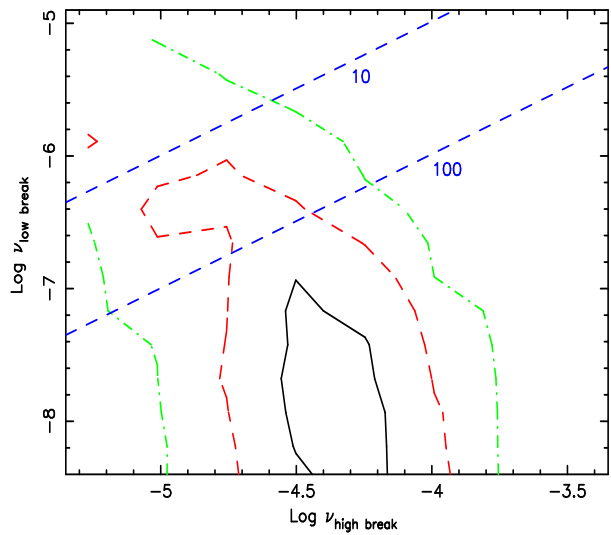}
\caption{Confidence contours for the high and possible low break
frequencies. The 68 percent contour is solid, the 90 percent contour is dashed and 
the 99 percent contour is dot-dashed. The straight dashed lines, with labels
of 10 and 100 respectively, indicate those ratios between the high and
possible low break frequencies. The fits
assume a break below the low frequency break of 0 and
an intermediate slope between the low and high frequency breaks of
0.8. Both break frequencies were allowed to vary, as was the slope above
the upper break, which was measured at $1.95\pm0.3$. 
}
\label{fig:2break}
\end{figure}

We searched for a second, lower frequency break, using the \xte and
{\it XMM-Newton} 4-10 keV data.  We assumed a low state PSD model,
fixing the slope below the lower break at 0, and the slope between the
lower and upper breaks at -0.8, i.e. the value we measure in the same
frequency range assuming a high state model. We allowed the two break
frequencies and the slope above the upper break to vary. The best fit
frequency for the higher break and for the slope above the higher
break was the same as in our high state fit and the best fit for the
lower break frequency ($6 \times 10^{-9}$Hz) was almost at the end of
the fitting range. The fit probability was 43 percent. In
Fig.~\ref{fig:2break} we show the 68 percent, 90 percent and 99
percent confidence ranges for the two break frequencies.  As the
typical ratio of the two break frequencies for a low state GBH is
between 10 and 30, we plot, as dashed straight lines, the limits
corresponding to a ratios of 10 and of 100.  The best fit ratio is
over 1000, but a ratio of 100 is just within the edge of the 90
percent confidence region.

We repeated the above analysis after fixing the intermediate slope at
$-1.0$ and obtained almost identical frequency ratios although the high
break frequency was then about half a decade higher in frequency.

On the basis of its PSD, the above analysis strongly suggest that
MCG--6-30-15 is the analogue of a high state GBH, although a low state
cannot be entirely ruled out.

\section{Mass Determination: Absorption Line Velocity Dispersion}
\label{sec:veldisp}

As discussed in the Introduction, the mass of the black hole in \mcg6
is not well determined. In particular there have been no reverberation
mapping observations, and no measurement of the stellar velocity
dispersion, the two techniques widely regarded as giving the most
reliable measurement of black hole masses.  However a reliable
measurement of the mass in \mcg6 is important for any discussion of
mass/timescale scalings in AGN and for any discussions of whether AGN
are in low or high states. In this section we therefore present
stellar velocity dispersion measurements from which we estimate the
black hole mass. In subsequent sections 
(Sec~\ref{sec:linewidth} and Sec~\ref{sec:photoion})
we present secondary determinations of the black hole mass, based on
the width of the $\rm H \beta$ emission line and on photoionisation
calculations. We find that all these optically based mass determination
methods give consistent answers.

\subsection{Observations}
\label{sec:obs}

The observations of MCG--6-30-15 described here were taken by the
service programme of the 3.6m Anglo-Australian Telescope on 2002 June
5.  The RGO Spectrograph was used, with the 25cm camera plus the
1200R grating (blazed at 7500\,\AA).  The detector is an EEV2 CCD,
windowed to $600 \times 4096$ pixels for the science observations and
$150 \times 4096$ pixels for the standard stars.  A slit width of
$1''$ (0.15mm) was chosen, resulting in a spectrum of resolution R
$\sim 9500$ covering a wavelength range of 1000\,\AA.  The slit was
placed along the major axis of the galaxy, at a position angle of
$116^\circ$.

The measured widths of the Ca{\sc ii} triplet lines are used to
estimate the black hole mass.  At the redshift of \mcg6, $z=0.00775$, the
Ca{\sc ii} lines are shifted to $\lambda\lambda 8566, 8610$ and
$8731$\,\AA, and therefore the central wavelength was chosen to be
$8620$\,\AA.  The total wavelength coverage of 1000\,\AA, spanning the
wavelength range $8120 - 9120$\,\AA, allows accurate determination
of the continuum on either side of the lines, and therefore correct
measurement of the line widths.

Observations of late-type giant stars, which have very narrow lines,
and which dominate the stellar light of nearby galaxy bulges, were
performed in order to measure the instrumental broadening and provide
templates for the cross-correlation analysis.  The K0 {\sc iii}
standard stars HD\,117927 and HD\,118131 were observed, with the same
set-up and wavelength coverage as \mcg6, since the rest wavelength of
the Ca{\sc ii} triplet ($\lambda\lambda 8498, 8542, 8662$\AA) lies within the
wavelength range covered.

Dedicated flat field exposures were taken before and after the science
and standard star observations, for accurate de-fringing, and argon
and neon arc spectra were taken for the wavelength calibration.

\subsection{Data Reduction}

The raw data were reduced with {\sc IRAF} using standard bias
subtraction, flat fielding and cosmic ray removal techniques.  At the
red wavelengths observed here, the EEV2 CCD suffers from fringing at a
level of $\sim 5$ percent.  A comparison of the extracted science spectra
with equivalent spectra taken from the flat field frames showed that
the fringes were efficiently removed from the science frames, and did
not contribute any spurious features.  Curvature in the spatial
direction was rectified using bright night sky lines in the object
frames.  The narrow window used for the standard star observations
meant that in this case, rectification was unnecessary.  The continua
of \mcg6 ~and the standard stars were sufficiently bright that curvature
of the spectra in the dispersion direction could be traced and rectified.
The total exposure time for \mcg6 ~was 1 hr, broken into two 30min
integrations.  Wavelength calibration was performed using the arc
exposures bracketing each science frame.  This calibration was checked
by measuring the resulting wavelength of night sky emission lines, and
no further correction was required.

\begin{figure}
\psfig{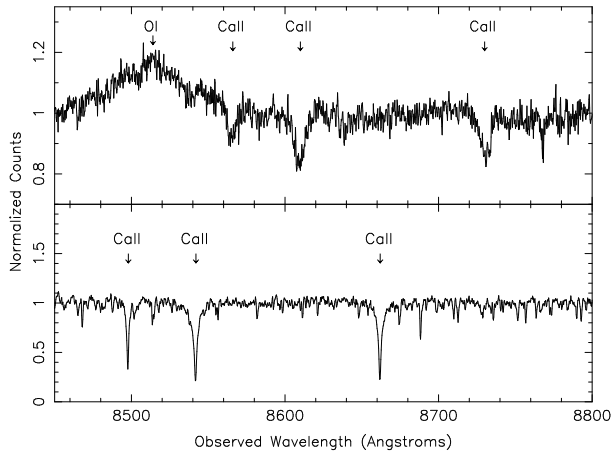}
\caption{Normalised spectra of \mcg6 (top panel) and a template star
(bottom), with the position of the calcium triplet feature
and broad O{\sc i} $\lambda 8446$ emission line marked.}
\label{fig:mcgspec}
\end{figure}

The nuclear spectrum of \mcg6 ~was extracted from the central 5.6 pixels
of the trace, which corresponds to a $2.4'' \times 1''$ aperture.
To be compatible with the $M_{\rm BH}-\sigma_{*}$ relation defined by
\cite{ferrarese00}, the length of this aperture was chosen to
correspond to $r_e/8$, where $r_e$ is the effective radius of the
galaxy, measured from the profile of the spectral data, compressed
along the dispersion direction.  We note however that the choice of
extraction aperture has only a very small effect on the results
obtained (Ho, private communication).  Sky subtraction was performed
using regions either side of the galaxy, 50 pixels ($21.5''$) wide, at
a distance of 50 pixels ($21.5''$) from the centre of the galaxy.  The
resulting spectrum is shown in Fig.~\ref{fig:mcgspec}, normalised by a
spline fit to the continuum to emphasize the emission and absorption
features.

In a similar analysis, \cite{filippenko03} detect Paschen emission
lines in the spectrum of NGC\,4395, adjacent to each of the calcium
absorption features.  For \mcg6, there is perhaps a suggestion of a
broad Paschen 15 line at $\sim 8615$\AA, but it is not conclusive.
Even if present, the strength of such a line is very small compared to
those detected in NGC\,4395, and as such, would not significantly
affect our results.

\subsection{Data Analysis}

The calcium absorption features in a galaxy spectrum are broadened
compared to those of an individual star due to the bulk motion of the
stars in the galaxy.  The extent of this Doppler broadening is related
to the mass of the central black hole, giving rise to the observed
tight relationship between the two quantities.  The region of the
galaxy spectrum of \mcg6 ~containing the calcium triplet is therefore
cross-correlated with the equivalent region of the K0{\sc iii} stellar
templates using the {\sc IRAF} task {\tt fxcor}.  The task fits a
spline function to the continuum of each input spectrum, and
cross-correlates the resulting features, giving as output a velocity
measurement, due to the redshift of \mcg6, and a velocity width, due
to the Doppler effect of the motion of stars in the galaxy caused by
the gravitational influence of the black hole.

The width of the cross-correlation peak function calculated by {\tt
fxcor} measures the combined Doppler and instrumental broadening,
therefore in order to ascertain the Doppler broadening alone, the
stellar templates themselves were broadened using a Gaussian of
various widths, and cross-correlated with the unbroadened templates.
The width of the Gaussian which gave rise to a cross-correlation peak
width equal to that of \mcg6 ~provides our measurement of the true
velocity dispersion of the galaxy.  We found that broadening the
stellar templates by 11 pixels gave the best match to the
cross-correlation peak width of \mcg6.  The velocity dispersion of the
spectra is $8.5 \, {\rm km \, s}^{-1} \, {\rm pix}^{-1}$, giving a
velocity dispersion for \mcg6 of $\sigma = 93.5 \, {\rm km \,
s}^{-1}$.

The statistical errors on this measurement are very small, and
therefore do not provide a realistic estimate of the true errors
involved.  \cite{ferrarese01} estimate that the systematic uncertainties
involved in making such measurements are of order 15 percent.
We therefore undertook a number of tests in order to
ascertain the likely error in our result.  First, since we have three
standard star observations (two of HD\,117927 and one of HD\,118131),
each spectrum was broadened in turn and cross-correlated with the two
unbroadened stellar templates to estimate the variation due to the use
of differing standards.  The results usually varied by less than one
percent, and never more than two percent.  Secondly, to assess the
possible impact of the proximity of the O{\sc i} line to the first
feature in the calcium triplet, we restrict the region of the spectra
used in the cross-correlation analysis to that containing only two of
the three calcium absorption lines.  By changing the wavelength range
over which the cross-correlation analysis is performed, a maximum
variation of $\pm 4$ percent in the peak width was measured.  Thirdly,
we vary the amount of smoothing of the stellar templates required to
match the result from \mcg6, and find that all the results of the
above tests are easily bracketed by taking a Gaussian width of
$11\pm1$ pixels, which is equivalent to $\sigma = 93.5 \pm 8.5 \, {\rm
km \, s}^{-1}$ ($\pm 9.1$ percent).  We therefore take this as our
conservative estimate of the error in our measurement, and note that
this is small compared to the uncertainty in the relationship between
this quantity and the inferred black hole mass, as described in the
following section.

\subsection{Black Hole Mass Estimate}

The largest source of error in estimating the mass of the black hole
from the width of the absorption lines is not uncertainty in the
measurement of the width but uncertainty in 
the $M_{\rm BH} - \sigma$ relation itself. The two main groups concerned
with deriving the relationship use slightly different observational
methods and fit slightly different relationships to the resultant data
\cite[see][for full discussions]{merritt01, tremaine02}. 
Recently \cite{greene04} have shown that AGN with very low black hole
masses (down to $\sim 10^{4}$\msun) fit reasonably well onto the
relationship given by \cite{tremaine02}, i.e.,

\[
M_{\rm BH} = 1.35^{+0.20}_{-0.18} \times 10^8 M_\odot
\left(\frac{\sigma}{200 \, {\rm km \,s}^{-1}}\right)^{4.02(\pm0.32)}.
\]

\noindent
Our observations then imply $M_{\rm BH}=6.3^{+3.0}_{-2.0} \times 10^6 M_\odot$.
As can be seen from Fig.~4 of \cite{greene04}, the spread in the
datapoints implies an uncertainty in the derived mass of at least
50 per cent.

An alternative version of this relationship has been given by 
\cite{merritt01}. Using the most recent version of this alternative 
relationship \citep{ferrarese02} i.e.

\[
M_{\rm BH} = 1.66(\pm0.32) \times 10^8 M_\odot \left(\frac{\sigma}{200 \,
{\rm km \,s}^{-1}}\right)^{4.58(\pm0.52)},
\]

\noindent
we derive 
$M_{\rm BH}=5.1^{+3.8}_{-2.4} \times 10^6 M_\odot$,
which is entirely consistent with the mass derived using the
relationship of \cite{tremaine02}.

\section{Mass Determination: Emission Line Width}
\label{sec:linewidth}

An alternative estimate for $M_{\rm BH}$
can be obtained using the empirical relationships 
found by \cite{kaspi00} 
in reverberation studies of nearby Seyfert~1 galaxies and quasars
between $M_{\rm BH}$, the velocity dispersion of the broad emission-line gas,
$V_{\rm FWHM},$\footnote{While one might
expect $V_{\rm FWHM}$ of the root mean square profile to better
represent the velocity dispersion of the variable part of the BLR, and
thus show a better correspondence with $R_{\rm BLR}$, in practice
there is little difference in the derived masses 
\citep{kaspi00}.}
and the effective size, $R_{\rm BLR}$, of the broad line emitting
region, i.e.: 

\[
M_{\rm BH} =1.464\times 10^{5} \left(\frac {R_{\rm BLR}}{\rm{lt-days}}\right)
\left(\frac{V_{\rm FWHM}}{10^{3}~{\rm km~s}^{-1}}\right)^2 M_{\odot} \,.
\]

\noindent $R_{\rm BLR}$ is determined from the measured delay between the
continuum and emission-line variations and is related to the galaxy
subtracted continuum luminosity at $\lambda$5100\AA\ by

\[
R_{\rm BLR}=32.9^{+2}_{-1.9}\left( \frac{\lambda L_{\lambda}(5100{\rm
\AA})}{10^{44}\,{\rm erg~s}^{-1}}
\right)^{0.7} \,.
\]

\begin{figure}
\psfig{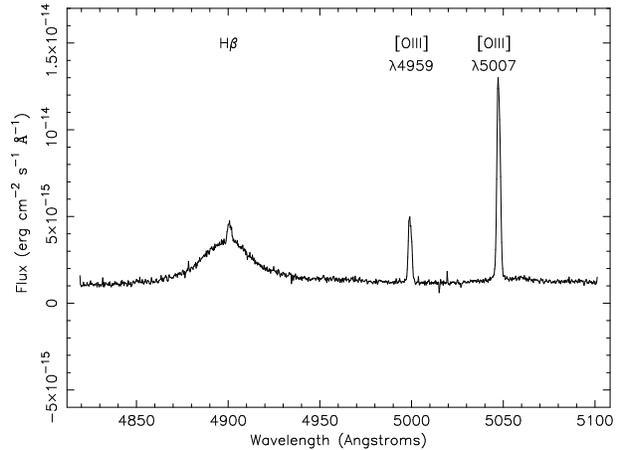}
\caption{{\it HST/STIS} spectrum of \mcg6, showing broad and narrow
$\rm H \beta$ components, and [O{\sc iii}] $\lambda\lambda \, 4959, 5007$\AA\,
emission lines.}
\label{fig:stis}
\end{figure}

\noindent \cite{reynolds97} discuss in detail the optical spectrum of
\mcg6 and, in their Fig. 4, derive an estimate of the intrinsic 
non-stellar optical
flux (ie $\nu F_{\nu}$, or $\lambda F_{\lambda}$) at 5100\AA of $\sim6
\times 10^{-11}$ ergs cm$^{-2}$ s$^{-1}$. Assuming the standard flat
cosmology of $H_{\rm 0}=75$~km~s$^{-1}$~Mpc$^{-1}$ and
$\Lambda_{matter}=0.3$, with $z=0.007749$, this flux equates to a
rest-frame luminosity of $\lambda
L_{\lambda}=7.2\times10^{42}$~erg~s$^{-1}$ and yields $R_{\rm
BLR}=5.2$~lt-day.

We note that there is considerable scatter in the relationship between 
$R_{\rm BLR}$ and $\lambda F_{\lambda}$ at 5100\AA.
\cite{vestergaard02} repeated the empirical study of
\cite{kaspi00} using the linear regression techniques of \cite{akritas96a}
and obtained a slightly different relationship. However the derived value of
$R_{\rm BLR}$, ie 5.36~lt-day is, within the errors, identical to 
the value derived using the relationship of \cite{kaspi00}.
\cite{vestergaard02} derives the relationship using
only broad line Seyfert 1 galaxies. The narrow line Seyfert galaxy,
NGC4051, is an outlier and is not used in deriving the relationship,
although it is used by \cite{kaspi00}.

Using archival {\it HST/STIS} data for \mcg6
(Fig.~\ref{fig:stis}) we have measured (rest frame) $V_{\rm
FWHM}=32.9\pm 1.9$\AA\ ($\equiv 2020\pm 120$~km~s$^{-1}$) which,
together with $R_{\rm BLR}$,
gives a virial mass for the black hole in \mcg6 of $3\times
10^{6}$\msun.

\section{Mass Determination: Photoionisation}
\label{sec:photoion}

The size of the BLR may also be found from photoionization
calculations, but with relatively large uncertainty. The calculated
BLR size, combined with the velocity at FWHM of the $\rm H \beta$\
line profile can then be used to provide an estimate for the virial
mass of the central black hole.  From a sample of 17 Seyfert~1
galaxies and 2 quasars, \cite{wandeletal99} found an approximately
linear relationship between photoionization mass estimates and
reverberation mass estimates, suggesting that photoionization
calculations might providing a route for mass determinations in
systems with only single epoch spectral observations. The empirical
relationship derived by \cite{wandeletal99} is:

\[
M_{\rm BH} = 2.8\times 10^{6}f\left( \frac{L_{\rm 44}}{ Un_{10}}\right) ^{1/2}
\left(\frac{V_{\rm FWHM}}{10^{3}~{\rm km~s}^{-1}}\right)^2 M_{\odot}
\, .
\]

\noindent Here $U$ is the ionization parameter (the dimensionless ratio of
photon to gas density), $n_{10}$ is the electron density in units of
$10^{10}$~cm$^{-3}$, and $f$ is the product
$f_{k}f_{L}^{1/2}\overline{E}^{-1/2}$, where $f_{k}$ is a factor
relating the effective velocity dispersion to the projected velocity
dispersion, $f_{L}$ relates the observed luminosity $L$ to the
ionizing luminosity, and $\overline{E}=\overline{E}/1~\rm{ryd}$.  If
we assume a BLR ionization parameter and gas density typical for
nearby Seyfert~1 galaxies ($U=0.1$, $n_{10}=10$), and a weighted
$f$-value $\sim1.45$ \cite{wandeletal99} we derive a black hole mass
of $4.5\times 10^{6}$\msun. 

\section{Discussion}
\label{sec:disc}

\subsection{Summary of Mass Determinations For MCG--6-30-15}

Although considerable systematic uncertainties and large assumptions
(eg in $n_{10}$ and $U$) are involved in their derivations, the
various optical measurements of the black hole mass in \mcg6,
including the revised estimate based on the bulge mass, are in
reasonable agreement. All values lie between $\sim3$ and $\sim6 \times
10^{6}$\msun. We cannot tell which measurement is correct and so, as a
working value, we take the middle of the range, ie $\sim4.5 \times
10^{6}$\msun and adopt an error equal to the spread in the
measurements (ie $3 \times 10^{6}$\msun).  We note that although there
are again large uncertainties in the mass derived from the PSD, the
most tightly constrained mass, ie that derived from a combination of
\xte and low energy {\it XMM-Newton} observations and assuming linear 
scaling of
break timescale with mass from Cyg X-1 in the high state, ie
$\sim2.9^{+1.8}_{-1.6} \times 10^{6}$\msun, is in good agreement with
the mass as determined by optical methods.

The average 2-10 keV X-ray flux of \mcg6 from our \xte monitoring is
$5.9 \times 10^{-11}$ \ecs, which corresponds to a luminosity of
$\sim7 \times 10^{42}$ ergs s$^{-1}$. For an assumed X-ray/bolometric
correction of 27 \citep{padovani88,elvis94}, the bolometric luminosity is $1.9
\times 10^{44}$ ergs s$^{-1}$, implying that, for a mass of $\sim4.5
\times 10^{6}$\msun, \mcg6 is radiating at $\sim40$ percent of its Eddington
luminosity. For narrow line Seyfert galaxies, the bolometric
correction should probably be less than 27 but probably still greater
than 10, so MCG--6-30-15 is almost certainly radiating at $>10$ percent 
of its Eddington luminosity. A low state interpretation of the PSD
which, although very unlikely, cannot be ruled out entirely, 
implies a black hole mass of $\sim 5 \times 10^{5}$\msun,
thus requiring a super-Eddington luminosity.

\subsection{X-ray Selected AGN and High State PSDs}

\begin{figure}
\psfig{figure=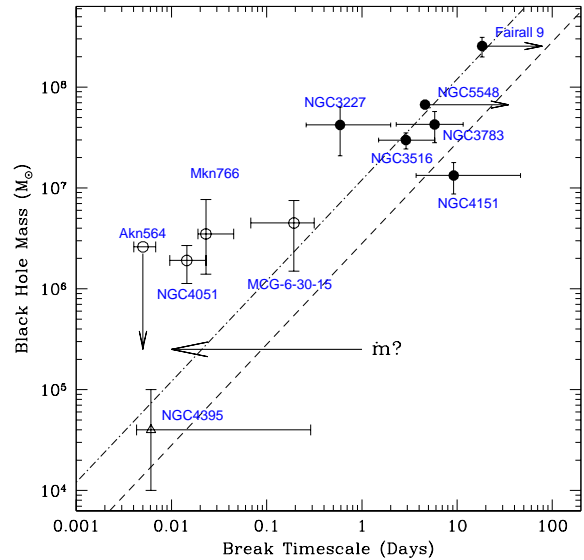,angle=0,width=8cm}
\caption{Black hole mass vs. PSD Break Timescale. Broad line Seyfert 
galaxies are shown as filled circles, narrow line Seyfert galaxies are
shown as open circles and NGC4395 is shown as an open triangle.  The
dashed line to the right represents linear scaling of break timescale
with mass from Cyg X-1 in its low state. The dot-dash line, on the
left, represents linear scaling of break timescale with mass from Cyg
X-1 in its high state.  NGC4395
is plotted as an open triangle (see text).
The arrow labelled with $\dot{m}$ indicates the way in which the
relationship of 
break 
timescale vs. mass might move with increasing accretion rate.
}
\label{fig:bhmcg6}
\end{figure}

Using our newly derived black hole mass, and slightly refined break
timescale, we plot (Fig.~\ref{fig:bhmcg6}) MCG--6-30-15 on a revised
version of the break timescale/black hole mass diagram which we
presented in M$\rm^{c}$Hardy et al. (2004). Although
bending, rather than sharply breaking, powerlaws fit the PSDs best, we 
still use timescales derived from sharp breaks in this diagram as we
do not yet have timescales derived from bending powerlaw fits to all
of the AGN. Like NGC4051,
\mcg6 also sits above the high state line. The break timescales are
taken from the compilation in \cite{mch04} with the addition of
NGC4395 from \cite{vaughan05}, with mass estimate from
\cite{filippenko03}. \cite{kraemer99} present UV and optical spectral
of NGC4395 and show broad, although
rather weak, wings to $H \beta$. They estimate 
FWHM$\sim$1500 km s$^{-1}$, but do not give an error. We are
therefore unsure whether to class it as a broad or narrow line Seyfert 
galaxy and so mark it by an open triangle in Fig.~\ref{fig:bhmcg6}.
We also include NGC3227 from Uttley and \mch (in preparation).  Uttley
and \mch also discuss NGC5506 in considerable detail but we do not
include it here as its mass is highly uncertain.

Where available (ie Fairall9, NGC3227, NGC3783,
NGC4051, NGC4151, NGC5548) we use reverberation masses from the
compilation of \cite{peterson04}. No reverberation masses are
available for the other AGN, ie the narrow line Seyfert 1s, and so
masses are derived from stellar velocity dispersion measurements.  For some
objects alternative masses are available from the emission line
width (eg $6.3 \times 10^{5}$\msun~for Mkn766 from Botte et al. 2005) but to
reduce the number of variables we restrict ourselves to the velocity
dispersion mass ($3.5 \times 10^{6}$\msun~for Mkn766, using the
calibration of Tremaine et al. 2002). Current work
\citep[e.g.][]{ferrarese01,peterson04}
shows that masses derived from reverberation mapping and velocity
dispersion are consistent. An exception is Akn564. As in some other
NLS1s, the Ca{\sc ii} triplet lines in Akn564 are in emission
\citep{vangroningen93}, rather than
absorption, and so cannot be used to determine the black hole mass.
In this case the width of the [O III] 5007
emission line is used as a substitute for the width of the stellar
absorption lines in order to estimate the black hole mass
\citep{botte04}. The width of the [O III] 5007
emission line does correlate, although with considerable scatter
\citep{boroson03}, with the width of the Ca{\sc ii} triple stellar
absorption lines. However it has been noted \citep{botte05} that
the [O III] lines tend to be wider than the Ca absorption lines. Therefore,
assuming that the Ca absorption lines represent the black hole mass
more accurately, the [O III] line will typically give an overestimate
of the black hole mass, hence the upper limit on Akn564 in
Fig.~\ref{fig:bhmcg6}.

Due to recalibration to better fit the $M_{\rm BH}-\sigma_{*}$
relationship, masses in the sample of \cite{peterson04} have generally
increased from the earlier estimates used in \cite{mch04}.  Also, the
black hole mass estimate for NGC4051 from \cite{peterson04} is higher
than the estimate from \cite{shemmer03} which we used in
\cite{mch04}. Here, for consistency, we use the black hole mass
values from \cite{peterson04} wherever available.
Thus all AGN from the  \cite{peterson04} sample have moved
further away from the low-state line and
towards the high state line.  With the exception of NGC4151, all AGN
now lie above the low state line.  Although the `broad line' Seyfert
galaxies are, in general, closer to the low state line than the
`narrow line' Seyfert 1s, it is possible that all AGN considered here
might be high state systems.  Thus although \mdot may well have an
important effect on determining the break timescale, the accretion
rate in all AGN studied here may be high enough to make them `high
state' systems.  Indeed \cite{peterson04} show that the average
accretion rate for the current sample of AGN is roughly 10 percent of the
Eddington rate, which exceeds the $\sim2$ percent rate which typifies the
transition to the high state in galactic X-ray binary systems
\citep{maccarone03a}. For completeness we also include the break
timescale of NGC3227 from Uttley and \mch (2005, in preparation).
NGC3227 is an interesting galaxy, having broad permitted lines,
a hard X-ray spectrum, and a high state PSD. 
It is discussed in detail by Uttley and \mch (2005, in
preparation). 

An additional indicator of the `state' of an accreting black hole is
given by the ratio of its X-ray to radio luminosity
\cite[e.g.][]{gallo03,fender01,fender03}.
For Galactic black hole systems, a high X-ray/radio ratio signifies a
high accretion rate and a `high' state system. A low X-ray/radio ratio 
signifies a low accretion rate and a jet-dominated `low' state system.
The core radio flux of MCG--6-30-15 is very low, $\sim 1$mJy at 5GHz
\citep{ulvestad84}, and very similar to that of NGC4051
(\mch et al, in preparation). The X-ray fluxes, and hence the
X-ray/radio ratios, of both galaxies are high, again
indicating `high' state systems.

\subsubsection{A Possible Selection Effect}

It is interesting to note that the few AGN 
with sufficiently good PSDs
to distinguish between high and low states (NGC4051, MCG--6-30-15,
NGC3227), all have high state PSDs.  The AGN which we, and others
\cite[e.g.][]{markowitz03a}, have been monitoring are mainly taken
from the X-ray bright AGN which are visible in the bright-flux limit,
all-sky, X-ray catalogues
\cite[e.g.][]{mcha5}. As X-ray flux rather than, eg, radio flux, is the only
selection criterion in such surveys, there is a strong selection effect
towards selecting high state AGN. 

AGN with low state X-ray PSDs are likely to be found amongst those AGN
which have relatively more luminous radio emission \cite[cf][]{merloni03}. As
X-ray emission will not be the only parameter on which we select such
AGN, we may expect them to be, in general, fainter in X-rays than
those studied presently. Such AGN should be present in, eg, the sample
of radio and X-ray bright objects selected from the {\it ROSAT} and
VLA FIRST all-sky catalogues \citep{brinkmann00}.

The observations from \xte presented here, and elsewhere
\cite[e.g.][]{markowitz03a,mch04}, demonstrate the tremendous
importance of long timescale (years/decades) X-ray monitoring
observations for our understanding of AGN. \xte has revolutionised our
understanding and opened up many new exciting areas of study, and
hopefully it will continue to operate for many years to come. However,
in the longer term, it is crucial that a sensitive ($\leq$mCrab in a
few hours) all sky X-ray monitor, with a long lifetime ($\sim$decade) is
launched to carry on this exciting work, otherwise this newly emerging
field will die with \xte.

\bibliographystyle{/home/imh/tex/mnras}
\bibliography{/home/imh/tex/imh}

\section*{Acknowledgments}

This work was supported by grant PPA/G/S/1999/00102 to IMcH from the
UK Particle Physics and Astronomy Research Council (PPARC).
IMcH also acknowledges the support of a PPARC Senior Research
Fellowship. We thank Raylee Stathakis and the Service Programme of the
Anglo Australian Telescope for carrying out the observations very efficiently.
We thank Brad Peterson, Mike Merrifield and Andy Newsam for useful
discussions.

\label{lastpage}

\bsp

\end{document}